\documentclass[aps,prl,twocolumn,showpacs,superscriptaddress, preprintnumbers,amsmath,amssymb]{revtex4}

\usepackage{graphicx}
\usepackage{epsfig}
\usepackage{dcolumn}
\usepackage{bm}
\usepackage{longtable}
\usepackage{color}

\begin{document}


\title{Anisotropic hysteretic Hall-effect and magnetic control of chiral domains in the chiral spin states of Pr$_2$Ir$_2$O$_7$}


\author{L.\ Balicas}
\affiliation{Institute for Solid State Physics, University of Tokyo, Kashiwa 277-8581, Japan}
\affiliation{National High Magnetic Field Laboratory, Florida
State University, Tallahassee-FL 32310, USA}
\author{S.\ Nakatsuji}
\affiliation{Institute for Solid State Physics, University of Tokyo, Kashiwa 277-8581, Japan}
\author{Y.\ Machida}
\affiliation{Institute for Solid State Physics, University of Tokyo, Kashiwa 277-8581, Japan}
\affiliation{Department of Physics, Tokyo Institute of Technology, Meguro 152-8551, Japan}
\author{S.\ Onoda}
\affiliation{Institute for Solid State Physics, University of Tokyo, Kashiwa 277-8581, Japan}
\affiliation{Condensed Matter Theory Laboratory, RIKEN, Wako 351-0198, Japan }


\date{\today}

\begin{abstract}
We uncover a strong anisotropy in both the anomalous Hall effect (AHE) and the magnetoresistance of the chiral spin states of Pr$_2$Ir$_2$O$_7$. The AHE appearing below 1.5 K at zero magnetic field shows hysteresis which is most pronounced for fields cycled along the [111] direction.  This hysteresis is compatible with the field-induced growth of domains composed by the 3-in 1-out spin states which remain coexisting with the 2-in 2-out spin ice manifold once the field is removed.  Only for fields applied along the [111] direction, we observe a large positive magnetoresistance and Shubnikov de Haas oscillations above a metamagnetic critical field. These observations suggest the reconstruction of the electronic structure of the conduction electrons by the field-induced spin-texture.
\end{abstract}


\maketitle

Electronic transport under the influence of a background spin texture \cite{anderson} opened fundamental themes in physics, such as giant/colossal magnetoresistive effects \cite{maekawa,tokura} and the anomalous Hall-effect (AHE) \cite{nagaosa}. These magnetotransport phenomena provided the conceptual basis for current research in spintronics and have so far been intensively investigated in ferromagnetic, transition metals, oxides, and semiconductors \cite{ohno}. The recent discovery of a chiral spin liquid in the metallic frustrated magnet Pr$_2$Ir$_2$O$_7$ \cite{nakatsuji, machida} through the AHE at zero magnetic field below 1.5 K, provides a unique laboratory for studying novel magnetotransport phenomena in the absence of magnetic dipole long range order.

The AHE has long been regarded as a probe for bulk uniform magnetization in ferromagnets \cite{nagaosa}. In conventional ferromagnets the elimination of magnetic domain walls with the application of an external magnetic field, leads to a finite remnant magnetization $(M)$ at zero-field and to a concomitant Hall resistivity $\rho_{xy}$. In Pr$_2$Ir$_2$O$_7$, within the experimental accuracy, hysteresis was observed only in $\rho_{xy}$ and not in $M$ above a partial spin-freezing temperature, $T_f$, which ranges from  0.12 to 0.3 K\cite{nakatsuji, machida}. The Pr $\langle111\rangle$  magnetic moments are responsible for most of the magnetic properties in this material, excepting for a 10 \% reduction in their value due to screening by the Ir conduction electrons.  These moments are believed to be subjected to strong geometrical frustration\cite{nakatsuji, machida} in analogy to spin ice systems \cite{ gingras}, leading to the suppression of magnetic dipole long-range order. These observations indicate the emergence of a chiral spin state where the time-reversal symmetry is broken on a macroscopic scale \cite{comment}. The order parameter is most likely a uniform component of the scalar spin chirality \cite{wen}
$\kappa_{ijl} = \overrightarrow{S}_i \cdot \overrightarrow{S}_j \times \overrightarrow{S}_l$, which is related to the solid angle $\Omega_{ijl}$ subtended by three local spins $S$ through  $\kappa_{ijl} = \sin \Omega_{ijl}$, and to a delocalized orbital current. In effect, the electron wave-function acquires a finite Berry-phase\cite{nagaosa, taguchi}, even when the spins remain fluctuating\cite{onoda1,metalidis}, which induces the AHE. Therefore, the observation of the AHE without magnetic dipole order points to an emergent chiral spin state\cite{machida}, with highly anisotropic magnetotransport properties in a cubic system, and opens the possibility of controlling chiral ordered domains by the applied magnetic field.

\begin{figure*}[t]
\begin{center}
\epsfig{file= 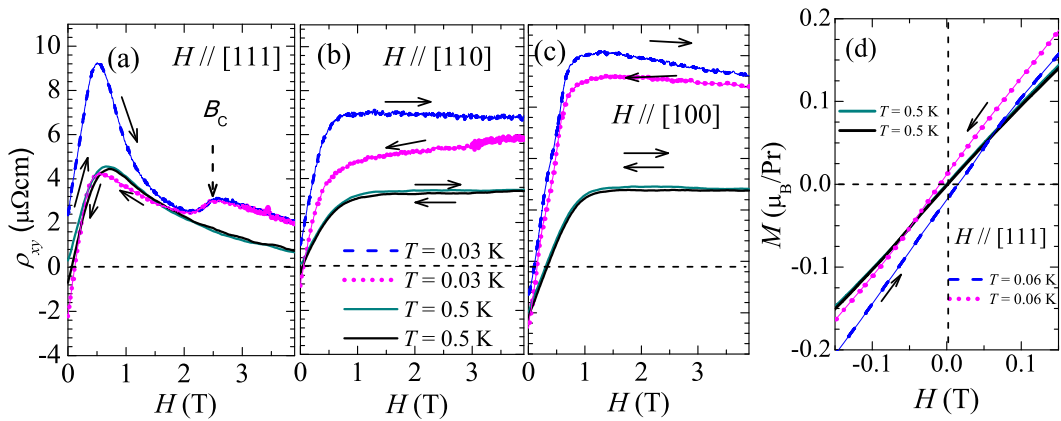, width = 14 cm}
\caption {(color online) Hall resistivity  $\rho_{xy} = R_{xy} \cdot t$, where $t$ is the sample thickness, for a Pr$_2$Ir$_2$O$_7$ single crystal, as a function of magnetic field $H$ along (a) [111], (b) [110], and (c) [100] and for $T = 30$ and 500  mK, respectively. These traces were measured for increasing (blue and clear blue traces) and decreasing (magenta and red traces) field scans. (d) Magnetization $M$ as a function of $H$, for increasing and decreasing field scans, and respectively for $T = 0.06$ and 0.5 K \cite{machida}. Notice the absence of hysteresis among the traces acquired at $T = 0.5$ K, but its presence in the Hall response, as shown in (a).}
\end{center}
\end{figure*}

Actually, the magnitude or even the emergence of the zero field Hall conductivity and its hysteresis depends on the application of an external magnetic field as well as on its strength \cite{machida}. This indicates that the field is linearly coupled to a particular chiral spin-texture, and thus might produce a strongly anisotropic response under field. But the chiral texture would have to survive the suppression of the field despite the fluctuating nature of the Pr spins in the spin-liquid phase. In order to unveil such unique physical response, here we performed a detailed study on the dependence of the AHE on temperature, time, field strength and its orientation. We reveal that the most pronounced contribution to the zero-field Hall-resistivity is observed when the Ir $5d$ conduction -electrons explore the Kagome layers perpendicular to the [111] axis. This contrasts sharply with conventional ferromagnets like Fe, Ni, and Co where the Hall resistivity is nearly isotropic \cite{nagaosa}. It suggests that the Kagome plane provides a basal plane for the chiral spin-texture. This is naturally understood if this texture incorporates through quantum fluctuations, the ``3-in 1-out"/``1-in 3-out" configurations for the four Pr moments in each tetrahedron, although being dominated by the ``2-in 2-out" spin-ice configurations \cite{hertog,bramwell1}.
In addition, our high field measurements have found a large positive longitudinal magnetoresistance and Shubnikov de Haas (SdH) oscillations only for fields applied along the [111] direction, and above a metamagnetic transition field. This suggests the reconstruction of the electronic structure of the Ir conduction electrons by the 3-in 1-out configuration induced by the external field.

We performed Hall-effect and resistivity measurements in oriented single crystals of Pr$_2$Ir$_2$O$_7$, grown by the method described in Ref. \cite{millican}, by using standard AC transport techniques. For the Hall measurements, samples were polished into a well defined bar-shaped geometry and contacts were attached in either a 4 or a 6 contacts Hall-geometry. Current was injected along the [110] crystallographic direction perpendicular to the magnetic field. Measurements were performed either in i) a $^3$He refrigerator, or in ii) a dilution refrigerator both coupled to 9 T superconducting solenoid, or in iii) a dilution refrigerator coupled to a 35 T resistive magnet. The Supplementary Material attached with this article shows the raw Hall resistance as a function of magnetic field for T = 0.5 K, exemplifying how the Hall signal and the magnetoresistance signal are evaluated, and that hysteresis is observed only in the Hall response.

Let us start with the chiral spin state at $T = 0.5$ K, which is well above $T_f$. To analyze the anisotropy in the Hall response of Pr$_2$Ir$_2$O$_7$, we show $\rho_{xy}(H) = R_{xy} \cdot t$ 
(where $R_{xy}$ is the Hall resistance normalized by the field and $t$ is the sample thickness) measured with increasing and decreasing field along [111], [110], and [100] in Figs. 1 (a), (b), and (c), respectively. Notice the marked hysteresis seen around zero field. The increasing-field trace starts from a constant and positive Hall resistivity value at zero-field, while the decreasing field sweep tends asymptotically to the same magnitude at zero field but has the opposite sign. The hysteresis as observed here in  $\rho_{xy}(H)$ is virtually absent in  $\rho_{xx}(H)$. While the hysteresis and the remnant Hall resistivity are most pronounced for fields along the [111] direction which stabilizes 3-in 1-out or 1-in 3-out configurations,  they are appreciably reduced for fields along the [100] and [110] directions which favor 2-in 2-out configurations. For fields along the [111] direction the hysteresis can be closed by a moderate field strength ($\leq 2.6$ T). We stress that magnetization measurements, shown in Ref. \cite{machida}, and also included here in Fig. 1 (d), indicate that neither conventional magnetic order, such as ferromagnetism, nor glassy behavior is responsible for the behavior shown here for $T = 0.5$ K $> T_f$. These experimental findings apparently suggest the following scenario: the time-reversal breaking circulating current associated with the chiral spin state at 0 T is dominantly sustained in the planes perpendicular to the [111] directions. The associated spin chirality is then coupled to and thus controlled by the magnetic field applied along the [111] directions. However, it cannot be fully controlled by fields applied along the [100] and [110] directions, since they allow four and two different (111) planes to create chirality domains, respectively.

\begin{figure}[t]
\begin{center}
\epsfig{file= 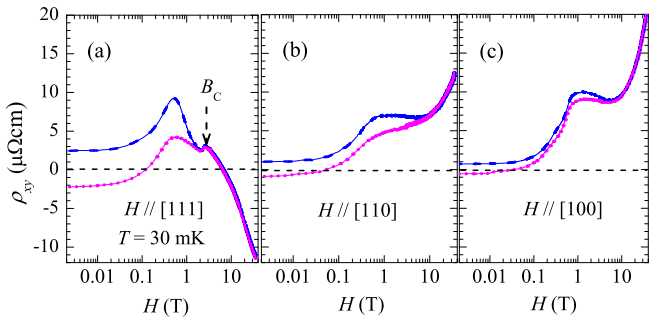, width = 8.6 cm}
\caption {(color online) Hall resistivity $\rho_{xy}$ measured at $T = 0.03$ K as a function of $H$ applied along  (a) [111], (b)  [110], and (c) [100] .  The kink at $B_c$ corresponds to a metamagnetic transition observed only for fields along the [111]-direction. A positive Hall response is found to increase and nearly saturate as the field increases, particularly for $H \|$ [110] and [100] directions, suggesting the progressive orientation and concomitant saturation of the population of tetrahedrons containing the 2-in 2-out spin configurations with a net moment along the field. As discussed in Ref. \cite{machida2} the 2-in 2-out state is characterized by a net positive chirality. For fields along the [111]-direction, the subsequent increase in $H$ leads to a change in the sign of  $\rho_{xy}$ which was attributed to the field-induced stabilization of tetrahedra containing the 3-in 1-out spin configuration with opposite chirality and concomitant negative contribution to the Hall response \cite{machida2}.  When $H$ is brought down to zero $\rho_{xy}(H = 0$ T) displays a negative asymptotic value indicating the presence of chiral domains composed of the 3-in 1-out spin configuration.}
\end{center}
\end{figure}

We have also explored the hysteresis in $\rho_{xy}$ at $T = 0.03$ K $< T_f$ . Figures 2 (a), (b), and (c) display $\rho_{xy}$ as a function of $H$ up to 35 T along the three principal axes. The hysteretic behavior is clearly more pronounced at the lowest temperatures with the asymptotic value $\rho_{xy} (H \rightarrow 0)$ being again largest for fields applied along [111]. $\rho_{xy}$ rapidly increases towards positive values with $H$ and remains nearly constant for $0.5 \leq H  \leq 10 $T. Above $\sim 10$ T, the Hall resistivity  becomes linear in field for $H \| $[110] and [100].  High fields along [111], according to the magnetization measurements \cite{nakatsuji, machida}, should stabilize the fully polarized ``3-in 1-out"/ ``1-in 3-out" state. In this high field limit, the Berry-phase calculation of the intrinsic AHE gives the Hall conductivity  $\sigma_{xy} = 12.6$  $\Omega^{-1}$cm$^{-1}$ within a simple model for the Ir $5d$ $t_{2g}$ conduction bands antiferromagnetically coupled to the localized Pr moments, as shown in Ref. \cite{machida}, leading to the correct sign of  $\rho_{xy}= - \sigma_{xy} /(\sigma_{xx}^2 + \sigma_{xy}^2)$. From Fig. 1,  $\sigma_{xy}$ is found to be field dependent reaching $\sim 21.5$ $\Omega^{-1}$cm$^{-1}$ at 35 T, which is remarkably close to the calculation.

\begin{figure}[t]
\begin{center}
\epsfig{file= 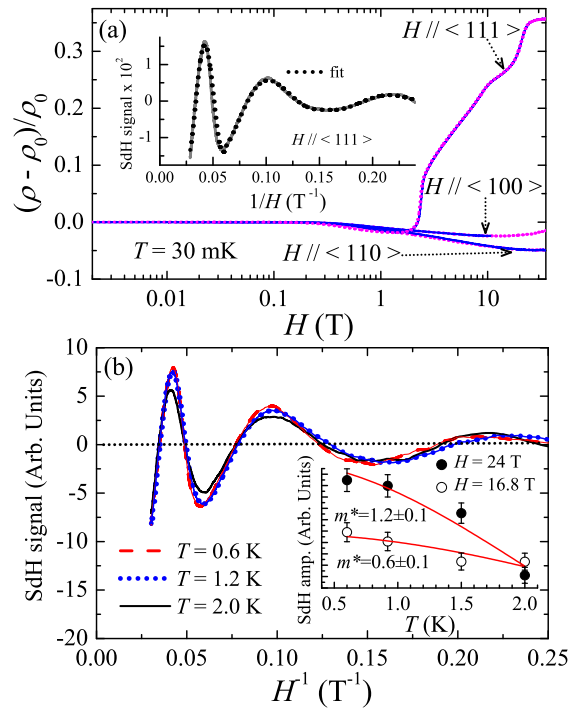, width = 7.4 cm}
\caption {(color online) (a) Magnetoresistance ratio $(\rho_{xx}(H)-\rho_{xx}(0))/\rho_{xx}(0)$ for all three principal axes at $T =0.03$ K and for both increasing and decreasing field sweeps. Inset: Oscillatory component in $\rho_{xx}$ obtained by subtracting the background signal after adjusting it to a fourth degree polynomial. Dotted line is a fit to a single Lifshitz-Kosevich oscillatory component with field-dependent frequency and Dingle temperature. For $H \geq 23$ T the system has reached its quantum limit and no Shubnikov de Haas oscillations are observed in fields up to 35 T. (b) Oscillatory component for the [111] direction superimposed into $\rho_{xx}$, as a function of $H^{-1}$ and for several temperatures. Red lines are fits to the Lifshitz-Kosevich formulae $x/ \sinh x$, where $x = 14.69 m^{\star} T/H$, for two values of magnetic field, i.e. 16.8 and 24 T, and from which we extract the respective effective masses $m^{\star}$. Both the geometry of the Fermi surface and the effective mass are field-dependent.}
\end{center}
\end{figure}

Figure 3 (a) shows the magnetoresistivity ratio as a function of field along three principal axes at $T =0.03$ K. When $H$ is applied along the [100] and [110] directions, the magnetoresistivity is small and negative,  suggesting the suppression of scattering by local Pr moments as they are oriented along the field. In contrast, for fields along the [111]-direction one observes a pronounced positive increase in magnetoresistivity above the metamagnetic critical field $B_c \cong  2.3$ T. It is followed by an oscillatory structure, shown in the inset. This oscillatory component is not periodic in $H$ nor in $H^{-1}$, although its nearly exponential growth as a function of $H$ indicates that it corresponds to SdH oscillations. In fact, one can fit it to a single Lifshitz-Kosevich oscillatory term containing a single field-dependent frequency (dotted line) implying that the associated Fermi-surface cross-sectional area is field-dependent. This probably results from the effect of the Zeeman splitting on a very small Fermi-surface whose frequency $F < 10$ T at low-fields, increases up to $\sim 22$ T at higher fields. As seen in Fig. 3 (b) the oscillations are still very well-defined at much higher temperatures. The temperature dependence of their amplitude indicates quite light effective masses which are also field-dependent, see inset of Fig. 3 (b). SdH oscillations at low $T$s imply a well-defined Fermi-surface and a relatively long mean free path for Pr$_2$Ir$_2$O$_7$ ranging from 255 {\AA} at $\sim$10 T to 850 {\AA} at $\sim$25 T and a carrier density ranging from $1.0 \times 10^{23}$ m$^{-3}$ to $5.7 \times 10^{23}$ m$^{-3}$, or respectively $5.63 \times 10^{-5}$ and $3.2 \times 10^{-4}$ carriers per Pr. This very small number suggests that we did not detect the heaviest parts of the Fermi surface. The oscillations are seen only for fields applied along the [111]-direction and above $B_c$ where the 3-in 1-out spin-configuration becomes dominant \cite{machida2}. This indicates that the field-induced change, from a state containing all possible degenerate 2-in 2-out spin configurations to a state dominated by a uniform  3-in 1-out spin configuration, can actually lead to the reconstruction of the Fermi surface.

\begin{figure}[t]
\begin{center}
\epsfig{file= 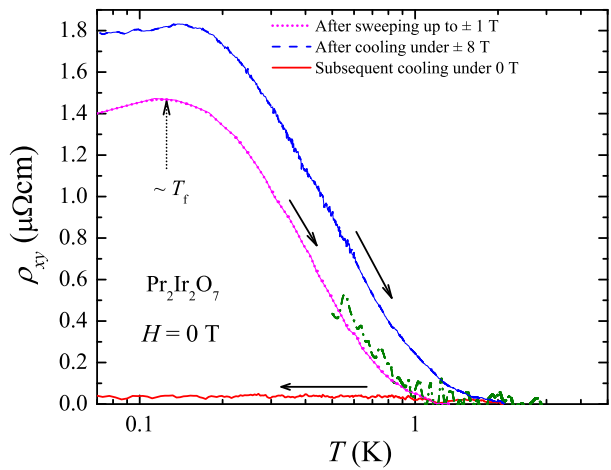, width = 8 cm}
\caption {(color online) Zero-field Hall resistivity $\rho_{xy}$ as a function of increasing $T$ measured after the field was swept to $ \pm 1$ T (or without crossing $B_c$) at respectively $T = 0.075$ K (magenta) and at $ 0.5$ K (green). The raw signal as a function of temperature was measured after the field was slowly ramped down to zero at a rate of 1 T/hr. The blue trace correspond to $\rho_{xy}$ measured with increasing $T$ after the sample was cooled from 2 K to 0.05 K under $\pm 8$ T, respectively.  In each case, $\rho_{xy}$ was obtained by subtracting one trace from the other in order to suppress the resistive component. In a similar manner the red trace was obtained from both curves obtained after the subsequent cool-downs at zero-field.}
\end{center}
\end{figure}

In Fig. 4, we show the temperature dependence of $\rho_{xy} (H = 0)$, as defined previously, but obtained in two different ways. The blue curve was obtained after cooling down the sample from 2 to 0.05 K under $\pm 8$ T. As seen, $\rho_{xy}$ is a smoothly decaying function of the temperature and is fully suppressed by $T \sim 2$ K. The subsequent cool-downs from 2 K to base temperature under zero-field still lead to a very small but finite spontaneous Hall signal (red trace). The magenta curve was obtained after ramping the field down to zero from $\pm 1$ T (or without crossing $B_c$). $\rho_{xy}$ shows a similarly decaying temperature dependence, but in the present case it completely disappears at $T \sim 1.25$ K. While the green curve was obtained after ramping down the field from $\pm 1$ T at $T = 0.5$ K $> T_f$. The green trace nearly overlaps the magenta one. Thus, the $T$ dependence of the hysteresis depends on the original strength of the field used to induce it but very little on the temperature chosen for the field scan.

In magnetic systems, hysteresis is generated by the field-induced alignment of domains with the temperature playing the opposite role by creating randomly oriented domains. In Pr$_2$Ir$_2$O$_7$, the field seems to favor the growth of domains with a particular chiral spin texture, leading to a finite Hall signal once the field is brought down to zero, with larger domains requiring larger $T$s to randomly orient them. These domains are seemingly very stable in time, since at the lowest $T$s the amplitude of the zero-field Hall signal does not decay within a few days. Since the largest hysteresis and concomitant AHE signal in Pr$_2$Ir$_2$O$_7$ are observed for fields applied along the [111]-direction, one is led to conclude that these field-induced domains are composed by the local 3-in 1-out spin configuration, embedded in a large ``matrix" containing 2-in 2-out spin states which remain fluctuating for $T > T_f$.
In this regard, as a recent theory on Pr pyrochlores predicts \cite{onoda}, ``3-in 1-out" and ``1-in 3-out" configurations, which would act as Dirac monopoles and form Dirac strings in the dipolar spin ice \cite{castelnovo,morris,bramwell,fennell}, should proliferate in the ground state dominated by ``2-in 2-out" configurations, because of the quantum nature of the Pr moments. Our observations pose the intriguing possibility that magnetic monopoles, which are classical defects in the case of dipolar spin ice, might acquire a quantum kinetics and finite quantum-mechanical average in the zero-field chiral spin-liquid state of Pr$_2$Ir$_2$O$_7$.

The authors acknowledge discussions with M. Tokunaga,T. Tomita, D. F. Haldane, S. L. Sondhi and K. Yang. LB acknowledges ISSP at the University of Tokyo for its hospitality and financial support. This work is partially supported by Grants-in-Aid (Nos. 21684019 and 21740275) from JSPS, by Grants-in-Aid (No. 19052003 and No. 19052006) and (No. 20102007) from MEXT-Japan and by Toray Science Foundation. The NHMFL is supported by NSF through NSFDMR-0084173 and the State of Florida. LB is supported by DOE-BES through award DE-SC0002613 .


\begin{thebibliography}{}
\bibitem{comment} The onset of the spontaneous Hall conductivity at $\sim 1.5$ K requires
an emergent macroscopic order parameter that breaks the time-reversal
symmetry. Thus, one can exclude any possible role of the partial, random freezing of the Pr moments as seen at $T_f$.
\end{thebibliography}
\end{document}